      [1999/12/01 v1.4c Il Nuovo Cimento]
\documentclass{cimento}

\usepackage{graphicx}  

\title{The evolution of the mass-metallicity relation at z$>$3}
\author{R.~Maiolino\from{oar}\ETC,
T.~Nagao\from{naoj},
A.~Grazian\from{oar},
F.~Cocchia\from{oar},
A.~Marconi\from{unifi},
F.~Mannucci\from{ira},
A.~Cimatti\from{unibo},
A.~Pipino\from{uniox},
S.~Ballero\from{units},
A.~Fontana\from{oar},
G.L.~Granato\from{oap},
F.~Matteucci\from{units},
G.~Pastorini\from{unifi},
L.~Pentericci\from{oar},
G.~Risaliti\from{oaa},
M.~Salvati\from{oaa},
        \atque
L.~Silva\from{oat}}
\instlist{\inst{oar} INAF - Osservatorio Astronomico di Roma
          \inst{naoj} National Astronomical Observatory of Japan
          \inst{unifi} Universit\`a di Firenze, Dipartimento di Astronomia 
          \inst{ira} INAF - Istituto di Radioastronomia 
          \inst{oaa} INAF - Osservatorio Astrofisico di Arcetri
          \inst{unibo} Universit\`a di Bologna, Dipartimento di Astronomia
          \inst{uniox} Astrophysics, University of Oxford
          \inst{units} Universit\`a di Trieste, Dipartimento di Astronomia
          \inst{oap} INAF - Osservatorio Astronomico di Padova
          \inst{oat} INAF - Osservatorio Astronomico di Trieste
                          }
\PACSes{\PACSit{98.62.Ai,98.62.Bj}{}}
\begin{document}

\maketitle

\begin{abstract}
We present preliminary results of an ESO-VLT large programme (AMAZE) aimed at
   determining the evolution of the mass-metallicity relation
   at z$>$3 by means of deep near-IR spectroscopy.
   Gas metallicities and stellar masses are measured for
   an initial sample of nine star forming galaxies at z$\sim$3.3.
   When compared with previous
   surveys, the mass-metallicity relation inferred at z$\sim$3.3
   shows an evolution significantly stronger than observed at lower redshifts.
   There are also some indications that the
   metallicity evolution of low mass galaxies is stronger
   relative to high mass systems, an
   effect which can be considered
   as the chemical version of the galaxy downsizing.
   The mass-metallicity relation observed at z$\sim$3.3 is difficult
   to reconcile with the predictions of some hierarchical evolutionary models.
   We shortly discuss the possible implications of such discrepancies.
\end{abstract}

\section{Introduction}

The correlation between galaxy mass and metallicity has been known for a long
time \cite{lequeux79}. Thanks to the SDSS survey the mass-metallicity relation
has been recently confirmed and refined with a sample of more than 50,000
star forming galaxies \cite{tremonti04}.
The origin of this relation is ascribed to
various possible processes. One possibility is that outflows, originated by the
starburst winds, are responsible for ejecting enriched gas out of their host
galaxies.
In low-mass galaxies outflows may exceed the escape velocity, so that
freshly produced metals are lost into the intergalactic medium, therefore
yielding a lower effective enrichment; instead, the deeper gravitational
potential of massive galaxies is more effective in retaining metals, yielding
higher enrichment \cite{tremonti04}. Another possibility is that low mass
systems are little evolved: they have still
to convert most of their gas into stars, and therefore have little enriched
their ISM yet; instead, massive systems have converted most of their
gas into stars, therefore reaching maturity from the chemical point of view.
The latter scenario is commonly referred to as ``downsizing''. Finally, it has
been proposed that the IMF may change depending on the level of star formation,
so that the effective yield of metals is higher during the evolution
of galaxies with larger, final stellar masses \cite{koppen07}.

The relative role of these different processes in shaping the mass-metallicity
relation is debated. However, it is likely that each of them contributes at
least to some extent, since observational evidence has been found for each of
these processes. Each of these factors (outflows/feedback, downsizing, IMF)
has profound implications on the evolution of galaxies. Therefore, it is clear
that the mass-metallicity relation contains a wealth of information useful
to constrain models of galaxy formation and evolution.
Indeed, any model of galaxy evolution is now required to
match the mass--metallicity relation observed locally
\cite{kobayashi07,brooks07,derossi07}. However, different models predict
different evolutionary patterns of the mass-metallicity relation as a function
of redshift, and observational data are required to test and discriminate among
them.
Observational constraints of the mass-metallicity relation
have been obtained up to z$\sim$2.2 thanks to various deep surveys
(e.g. \cite{savaglio05,liang06,erb06}). However, at z$>$3, which
is a crucial redshift
range as we shall see in the following, the mass-metallicity relation has
been little explored yet.

\section{The AMAZE project}

\begin{figure}[!t]
\includegraphics[width=13.5truecm]{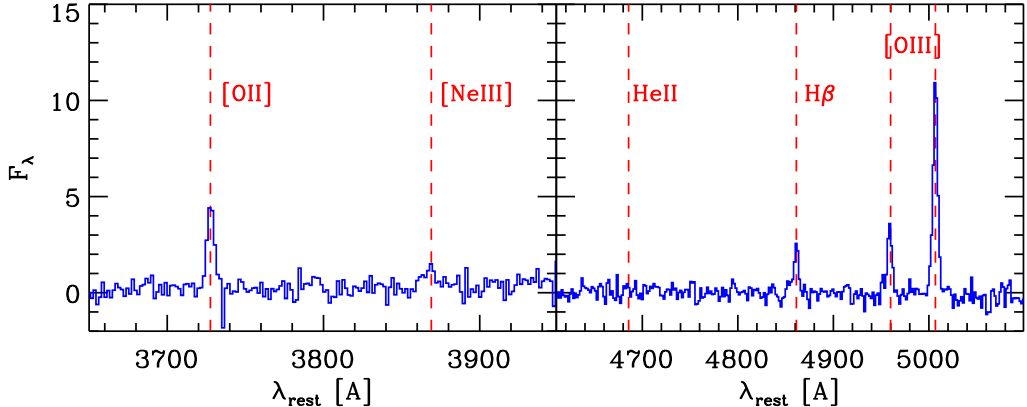}
\caption{Rest frame stacked spectrum of the first nine sources at 3$<$z$<$3.7
observed within the AMAZE programme. [OII]3727, [NeIII]3869, H$\beta$,
[OIII]4959, [OIII]5007 are the nebular lines used to determine the gas
metallicity. The expected location of HeII (4686), typically
observed in AGNs, is shown to highlight the absence
of AGN contribution.}
\label{fig_spec}
\end{figure}

We have started a project (AMAZE, Assessing the Mass-Abundance
redshift [Z] Evolution) specifically aimed at determining the
mass-metallicity relation at z$>$3. This is an ESO large programme
that has been awarded 180 hours of observations with SINFONI, the VLT
near-IR integral field spectrometer. The goal is to obtain near-IR spectra
of a sample of about 30 galaxies at 3$<$z$<$5. At these redshifts the near-IR
spectra allow us to measure the fluxes of the nebular emission lines
[OII]3727, [NeIII]3869, H$\beta$, [OIII]4959, [OIII]5007, whose relative
ratios can be used to constrain the gas metallicity \cite{nagao06}. In
particular, by combining all of the diagnostics that can be inferred
from these lines \cite{nagao06}, we can estimate
the gas metallicity by also accounting for the effect of possible
dust reddening, and
we can also control possible variations of the excitation conditions of the
gas (details of this procedure are given in \cite{maiolino07}).

The sample has been selected among Lyman Break Galaxies for which Spitzer-IRAC
data are available, which are required to obtain a good determination of
the stellar mass $\rm M_*$ at z$>$3 (where the 
rest-frame near-IR stellar light is redshifted to
$\rm \lambda > 3.5 \mu m$). AGN contamination must be absolutely avoided
(since it would affect the line ratios), therefore
we also required that the galaxies
have hard X-ray and Spitzer-MIPS data: hard X-rays allow us to identify the
presence of Compton
thin AGNs, while 24$\mu$m data allow us to identify even Compton thick
AGNs \cite{fiore07,daddi07}.

The observing programme is currently in progress. Here we summarize preliminary
results from the first 9 sources at 3$<$z$<$3.7 for which spectra have been
obtained and reduced (a more detailed discussion of this preliminary set of
data is given in \cite{maiolino07}). Fig.~\ref{fig_spec} shows the rest frame,
stacked spectrum of these 9 sources, where the nebular lines used to constrain
the gas metallicity are indicated. The expected location of HeII (4686),
typically observed in AGNs, is shown to highlight the absence
of any AGN contribution.

\section{The mass-metallicity relation at z$\sim$3.3}

\begin{figure}[!t]
\includegraphics[width=13.5truecm]{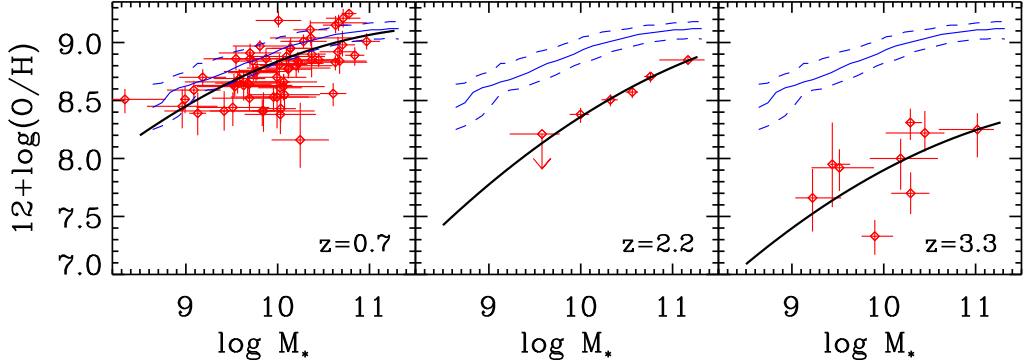}
\caption{Mass-metallicity relation observed at various redshifts (red points
with errorbars) compared
with the local relation \cite{tremonti04} (thin,
blue solid and dashed lines).
At z$\sim$0.7 (leftmost panel) we use the data from
\cite{savaglio05}, recalibrated with our relations. At z$\sim$2.2 (central panel)
we use the
data from \cite{erb06} (in this case the data are from stacked spectra and {\it not}
from individual galaxies), also recalibrated with our relations. At z$\sim$3.3
(rightmost panel) we show the preliminary results for the first nine galaxies in our
AMAZE program. Black, solid lines show quadratic fits to the relations at each
redshift.}
\label{fig_mzobs}
\end{figure}

Fig.~\ref{fig_mzobs} (rightmost panel)
shows the inferred mass-metallicity relation at z$\sim$3.3
(red points with errorbars) compared with the local mass-metallicity relation
(\cite{tremonti04}, blue thin lines, adapted to account for the IMF
adopted by us). The other panels show the
mass-metallicity relation at lower redshifts inferred by using results of
previous works \cite{savaglio05,erb06}, and where metallicities have
been re-determined by using the same
set of intercalibrated diagnostics adopted by us \cite{nagao06}
to ensures consistency between the various methods. Note that at z$\sim$2.2
the data points are obtained from stacked spectra \cite{erb06},
while at z$\sim$0.7 and at z$\sim$3.3 metallicities are inferred from the
spectra of individual galaxies.
The black, thick solid lines indicate quadratic fits to the data at each epoch.

\begin{figure}[!t]
\includegraphics[width=8truecm]{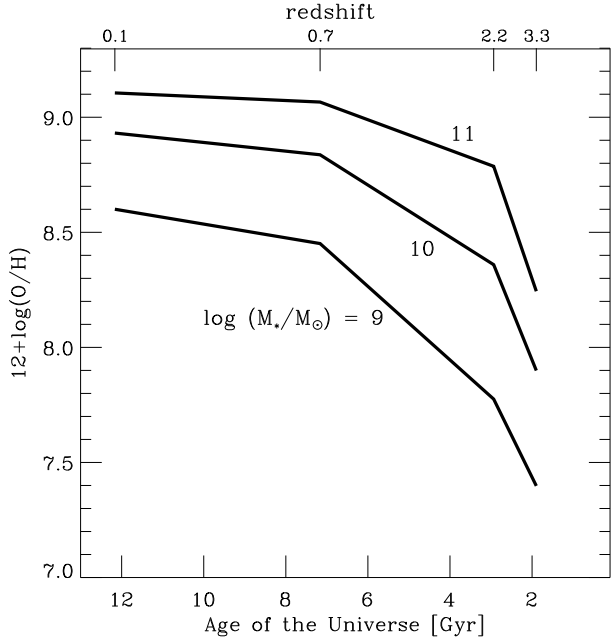}
\caption{Inferred, average metallicity evolution as a function of the age
of the universe, for different stellar masses. Note the strong evolution in
time between z$=$2.2 and z$=$3.3.}
\label{fig_evol}
\end{figure}

The mass-metallicity relation evolves significantly from z$=$0 to z$=$2, but
only by a factor of about two in metallicity (at high masses), which is not 
a very strong evolution if one considers that this redshift range embraces
75\% of the age of the universe. Between z$=$2.2 and z$=$3.3 the average
metallicity decreases by another factor of $\sim$3, but here the temporal
evolution is much stronger, since the time elapsed within this redshift range is
much shorter. This effect 
is shown more clearly in Fig.~\ref{fig_evol}, where the
evolution of the average metallicity (for different stellar masses) is plot
as a function of the age of the universe. The rate of evolution is clearly
steeper at 2.2$<$z$<$3.3 than at z$<$2.2 (especially at high masses),
indicating that in the former redshift
range we are witnessing an epoch of major action in terms of star formation and
of metal enrichment of galaxies.

Figs.~\ref{fig_mzobs}--\ref{fig_evol} also suggest that the evolution rate
is not constant with mass. At low stellar masses the evolution is
stronger than in massive systems. This can be regarded as the chemical version
of the galaxy ``downsizing'': massive systems reach chemical maturity at higher
redshift, while low-mass systems chemically evolve
more slowly and over a period of time possibly extending to the present epoch.
However, confirming this effect requires more statistics at low masses, where
our preliminary mass-metallicity relation is still poorly populated (and
possibly more prone to selection effects than massive systems).

When interpreting Figs.~\ref{fig_mzobs}--\ref{fig_evol} it is important to
bear in mind that at different redshifts surveys are sampling different
populations of galaxies. Local star forming galaxies sampled by SDSS in
\cite{tremonti04} are mostly spirals with modest star formation rates, while
LBGs used to investigate the mass-metallicity relation at z$\sim$2--3 are
characterized by enhanced star formation (and will likely evolve into massive,
quiescent local galaxies). Therefore, the patterns shown in
Figs.~\ref{fig_mzobs}--\ref{fig_evol} should not be interpreted as the evolution
of individual galaxies, but as the evolution of the average
mass-metallicity relation of
galaxies representative (or which contribute significantly to) the density
of star formation at each epoch. Additional issues related to possible
selection effects are discussed in \cite{maiolino07}.

Within this context it is interesting to note in Fig.~\ref{fig_mzobs} the
existence of some galaxies at z$\sim$3.3 with stellar masses approaching
$\rm M_*\sim 10^{11}~M_{\odot}$ and metallicity $\rm \sim
0.4~Z_{\odot}$. According to some models
\cite{granato04,pipino06}, these systems are expected to evolve into very
massive ellipticals (approaching $\rm 10^{12}~M_{\odot}$) with
solar/super-solar metallicities.

\begin{figure}[!t]
\includegraphics[width=11.5truecm]{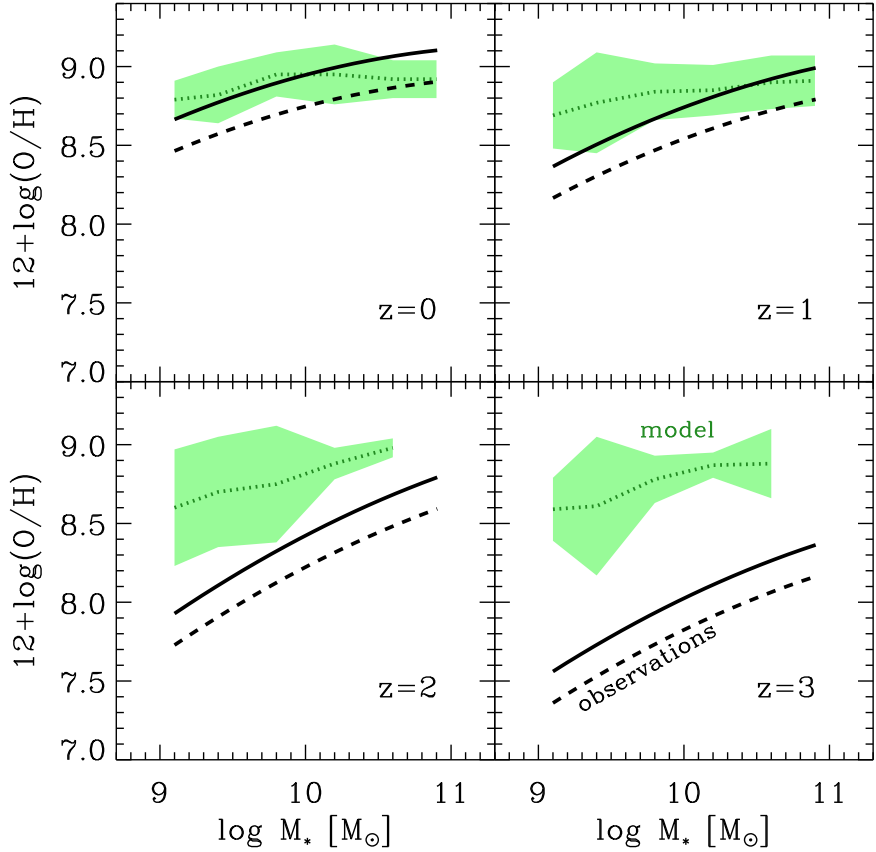}
\caption{Comparison of the evolution of the mass--metallicity relation
expected by the hierarchical model in \cite{derossi07} (green, dotted lines
and shaded areas) with the evolution measured in the observations (solid, black
lines). Observations have been interpolated linearly to match the redshifts
available for the theoretical model. The black, dashed lines show the observed
mass-metallicity relation shifted by 0.2 dex to account for a possible offset
of the metallicity scale \cite{erb06,derossi07}. Note the strong discrepancy
between the model and observations at z$=$3.}
\label{fig_mzmod}
\end{figure}

\section{Comparison with theoretical models}

Fig.~\ref{fig_mzmod} shows the comparison between the evolution of the
mass-metallicity relation inferred by the observations (black solid lines)
and the evolution expected by the simulations within a hierarchical scenario
as obtained by \cite{derossi07} (green dotted lines, where
shaded areas show the dispersion of the simulations). The dashed lines show the
observed mass-metallicity relation with an offset of 0.2~dex, to account for a
possible offset in the absolute metallicity scale (see
\cite{maiolino07,erb06} for discussions about this possible issue).
While at low redshift there is a fair agreement between model and observations,
at high redshift an increasing discrepancy emerges. In particular, at z$\sim$3.3
observations appears totally inconsistent with the prediction provided by the
theoretical model. Similar discrepancies are found by comparing observations
with the hierarchical simulations obtained by \cite{kobayashi07}.

The origin of the discrepancy between these hierarchical models and the
observations at high redshift is unclear. A possibility is that in these
hierarchical models most of the chemical evolution occurs rapidly in small
units, at low masses ($\rm M_* <10^9~M_{\odot}$). Therefore, according
to these models, at high redshift 
galaxies are mostly assembled with units that are already chemically
evolved, yielding a relatively flat relation and relatively high metallicity at
$\rm M_* >10^9~M_{\odot}$.
The lower metallicity found in the observations
at z$\sim$3.3 suggests instead that
galaxies at high redshift are made through the assembly of relatively unevolved
sub-units. In support of this scenario, we note that the alternative
hierarchical models presented in \cite{brooks07} match the mass-metallicity
relation at z$\sim$2.2 much better than the above mentioned models.
Indeed, one of the main features of the model in \cite{brooks07} is
that a strong
feedback prevents sub-units to evolve significantly before merging into more
massive systems. Firmer conclusions on the latter
class of models require a comparison
of the observations with the predictions at z$\sim$3.3, which are not
available in \cite{brooks07} yet.

A more detailed investigation will be possible thanks the completion
of the AMAZE project, within about one year.

\acknowledgments
We are grateful to S.~Savaglio for providing to us the electronic version of
the tables in her paper and for useful comments. We also thank A.~Brooks for
useful discussions.
This work was partially supported by INAF and by ASI.

\end{document}